# *Wedge Design*

A. R. L. Travis, Clare College, Cambridge CB2 1TL, UK. Email: arlt1@cam.ac.uk

Abstract

The space in front of a projector or camera can be folded into a light-guide by total-internal reflection if the guide is tapered like a wedge. This article explains how to calculate both flat and curved wedges by scaling a thin version designed using the principle that the product of thickness and the sin of ray angle is constant in a smoothly varying guide.

1. Introduction

A wedge-shaped light-guide can be designed so that multiple reflections fold up the space otherwise needed for rays to fan out from a projector or into a camera [1]. The approach has been considered for the projection of conventional and 3D images [2,3] and for imaging fingers that touch or hover in front of a display [4,5]. This article explains how to design a wedge.

2. The need for a slab

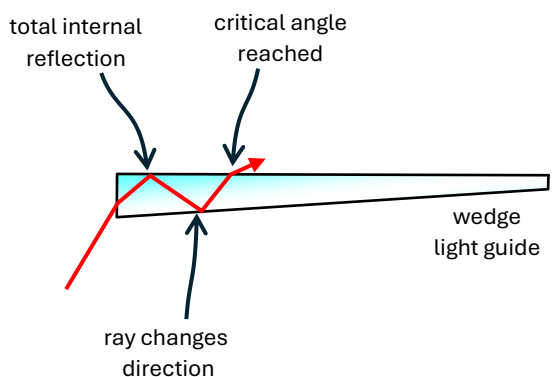

*Figure 1: The small angle between opposite surfaces of a wedge causes a guided ray eventually to emerge into air*

Shine a ray into the thick end of a transparent wedge and the ray will typically undergo total internal reflection off the wedge surface. The ray will proceed to the opposite surface that, because of the wedge shape, is at a small angle to the first so that if reflected, the angle of the ray with respect to the first surface will change slightly. If the angle of the ray was already close to the critical angle on first incidence, then this will be exceeded on subsequent incidence so that the ray will emerge into air as shown in figure 1.

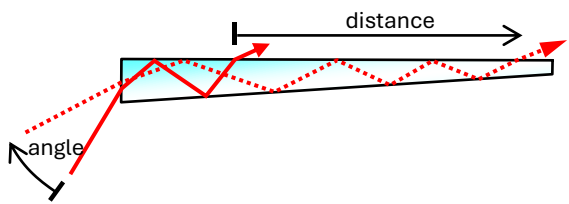

*Figure 2: The angle at which a ray is injected determines the distance that it travels before exiting the wedge*

It may be that the ray is launched at an angle very different from the critical angle in which case the ray will have to undergo many reflections before it reaches the critical angle and emerges into air. It follows that the angle at which the ray is launched determines the distance that it travels along the wedge before emerging into air as shown in figure 2. This property of converting launch angle into distance allows the wedge to expand the image from a projector even though the wedge is inherently slim.

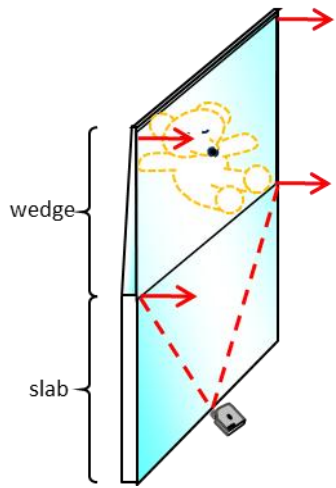

*Figure 3: Horizontal magnification takes place by fan-out in the elongated region that is called the slab*

The tapered profile expands the projected image along the axis of the wedge but it does not expand the image perpendicular to the wedge axis. It makes sense, therefore, to insert a slab waveguide of constant thickness between the projector and wedge so that rays can fan out to the width of the picture before rays begin to emerge from the system as shown in figure 3. Of course the slab and wedge are usually made as a single piece that is itself often called a wedge but for the purposes of analysis, it helps to define the wedge as beginning where the guide becomes thinner than at the system entrance.

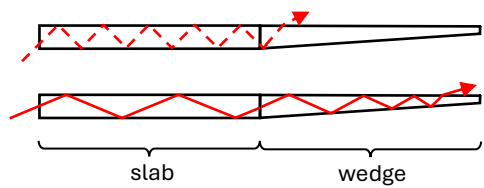

*Figure 4: A ray that undergoes many reflections in the slab will undergo few in the wedge and vice versa*

Rays launched directly into a wedge will leave the wedge on one or other side after an odd or even number of reflections. All rays must leave from just one side if we are to get an unbroken image without exotic launch conditions and here, the slab brings an extra benefit. Whereas the number of reflections in a wedge increases with reducing ray angle versus the wedge axis, the number of reflections in a slab diminishes as shown in figure 4. The sum of reflections in slab then wedge can be made approximately constant by choosing a slab with the appropriate length. Although the sum is approximately constant, it is not exactly so: how can we alter things so that it is?

3. <u>Design a thin wedge in which all rays undergo the same number of reflections before exit</u>

A well-established procedure in lens design [6] is to start with a thin design where the displacement of rays between the surfaces within a lens is made zero. The same approach works with wedges: it pays first to design a wedge as if it were very thin and only later model realistically thick systems [7].

Start by counting the number of reflections undergone by a ray at the critical angle in the slab. Whatever the shape of the wedge, this ray will emerge very soon after entering the wedge so it sets the number of reflections that all other rays must undergo before emerging from the wedge.

Now slightly reduce the ray launch angle relative to the guide axis, for example so that the ray undergoes one less pair of reflections in the slab as shown in figure 5 – notice that the critical angle also is defined relative to the guide axis. The ray must undergo a single pair of reflections in the wedge if it is to emerge after the same number of reflections as for the ray launched at the critical angle. We must therefore add a section of wedge long enough for this pair of reflections to take place whose length we calculate assuming the ray is at approximately the critical angle.

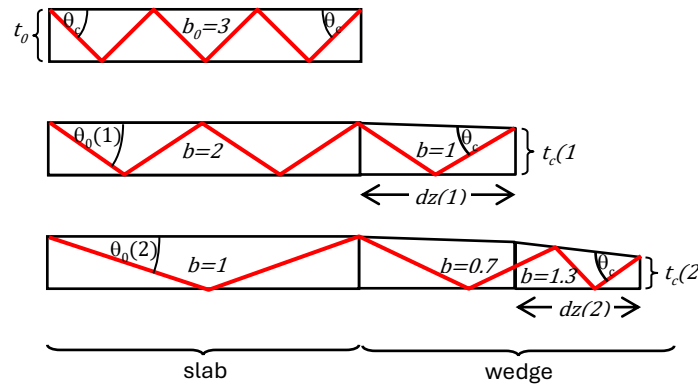

Figure 5: Add one section of the wedge at a time

The added section must of course be wedge shaped and an option is to find its slope by analysis of reflection angle. However, it is easier to use the principle that the product of guide thickness and ray angle is constant in a smoothly varying wave-guide (much like numerical aperture and the Lagrange invariant). This is proved below, see figure 6, and it tells us what must be the thickness at the end of the section of wedge added to the slab, provided that the system is very thin.

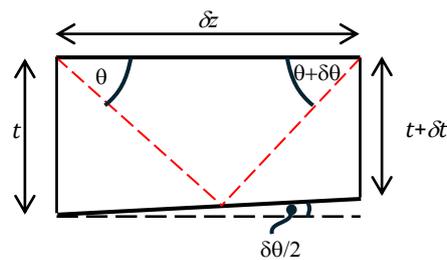

Figure 6: proof that $t \times \sin\theta$ is constant for a ray in a gradually varying wave-guide

| | | | |
|---|---|---|---|
| Geometry of bottom surface: | $\delta\theta/2$ | $= \dfrac{-\delta t}{\delta z}$ | (1) |
| Trigonometry of ray: | $\delta z$ | $= 2t/\tan\theta$ | (2) |
| Combining (1)&(2): | $-\dfrac{\delta t}{\partial \theta}$ | $= t/\tan\theta$ | (3) |
| Integrating (3): | $-\ln t$ | $= \int \cot d\theta + constant$ | (4) |
| | | $= \ln|\sin\theta| + constant$ | (5) |

Inverse log of (5):
$$\frac{1}{t} \propto \sin\theta \qquad (6)$$

Hence:
$$t \sin\theta = \text{constant} \qquad (7)$$

Again slightly reduce the ray launch angle relative to the guide axis, for example so that the array undergoes two pairs of reflections less in the slab than for a ray at the critical angle. This ray must undergo two reflections in the wedge so we need to count how many reflections it undergoes in the section of wedge that we have just added, then add another section of wedge of sufficient length. The thickness at the end of this second section is found once again by using the principle that the product of thickness and the sine of ray angle is constant.

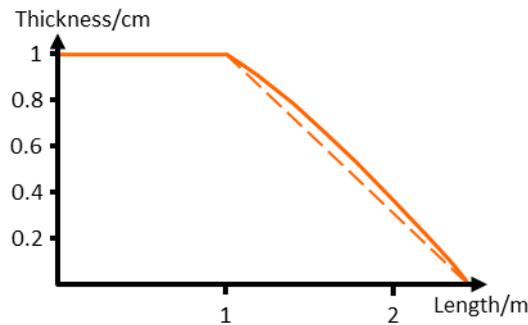

*Figure 7: The number of reflections to exit is the same for all rays if the wedge is slightly longer than the slab & slightly curved*

We repeat the procedure until the wedge has tapered to a tip and the result is shown in figure 7. The wedge is slightly bulged with respect to a flat and this bulge is not a regular curve of any kind. Note that we could have designed the wedge using conventional ray-tracing but that would require addressing the kink between wedge and slab before we are ready to do so.

4. <u>Scale up wedge thickness and add a bevel</u>

In principle, we are free to linearly scale the thickness and length of the slab plus wedge combination without changing its property that all rays undergo the same number of bounces. However as the system gets thicker, the kink - the sudden change in slope at the slab/wedge boundary - will cause the projected image to be banded. An effective solution is to fit a polynomial to the slab plus wedge – a sixth order polynomial typically has enough coefficients to follow large scale curvature while also smoothing out the kink. An alternative is to introduce a bulge into the slab whose final slope matches that at the wedge entry; this will be discussed later. However, introducing a bulge into the slab makes it thicker and the whole purpose of a wedge is to be thin.

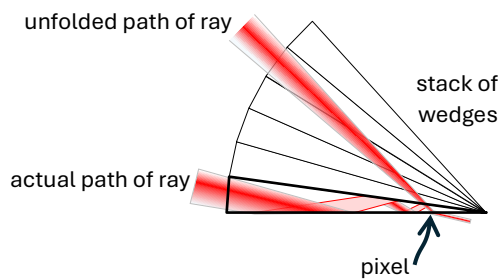

*Figure 8: The wedge must be thick enough to contain the entire ray despite aperture diffraction at the pixel*

We need to increase the wedge thickness so that the number of reflections is a finite integer but most often, the wish is that the wedge be as thin as possible, how thin can it be?

Note first that since a wedge is transparent, it can function in reverse as a slim periscope with the projector being replaced by a camera. Figure 8 shows such a system where collimated light illuminates a single pixel, i.e. an aperture, at the wedge surface. Light thereafter undergoes aperture diffraction within the wedge and we can find the effect of this aperture diffraction by tracing the ray as a straight line through multiple reflections of the wedge. The entire ray must emerge from the thick end if we are to resolve the pixel and the wedge must be thick enough to make this possible. It is therefore resolution that determines how thick the wedge must be. Geometry can give us a formula for figure 8 but our wedge profile will typically be more complicated than a simple pair of flats so ray-tracing is needed to verify what resolution is feasible. Nevertheless, the principle that thickness determines resolution applies both to cameras and projectors.

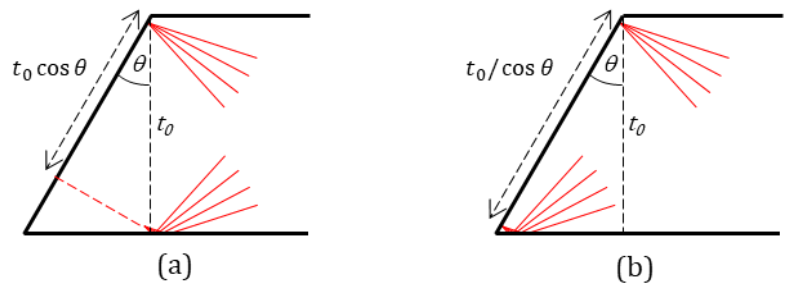

*Figure 8: Adding a bevel has the effect of (a) reducing entrance aperture if upper and lower claddings are the same (b) increasing it if they are somewhat different*

Once a finite wedge thickness has been chosen, it makes sense to bevel the input in order to facilitate the task for the projection/camera lens. It might seem that the bevel increases the entrance aperture and therefore the potential resolution of the system but that depends on what surrounds the wedge.

If both the bottom and top surfaces of the wedge are surrounded by the same material e.g. air, then infinitesimally thin wedges are symmetrical. This means that if a wedge is designed so that all rays entering the top surface are incident on the thick end in a downward-going direction, then all incident on the bottom surface are incident on the thick end in an upwards-going direction. Figure 8(a) shows how all rays that enter one surface of an ideal wedge will behave at its thick end and we see that the effective entrance aperture is less than the thickness of the wedge. The aperture need not be at the bevel surface and it may ease design of the projection lens if the aperture is pulled back from the bevel surface.

If, on the other hand, the wedge is asymmetrical, for example if one surface interfaces to air while the other interfaces to e.g. an aerogel of slightly higher index than air, we can have the situation of figure 8(b) where the entrance aperture is greater than wedge thickness. This is also possible with a slightly curved wedge where rays are more likely to emerge from the outer curved surface than the inner curved surface, even if both surfaces interface to the same cladding material.

5. Worked example

Suppose we want to design an acrylic wedge with a thickness of 5 mm and an image length of 50 mm. Run algorithm flatslab.m of appendix 1 in Octave or Matlab® and it will produce an array of thicknesses t and an array of distances z that define the slab and wedge. We see that the wedge has approximately twice the required length (it's best not to take it all the way to the tip) and is too thin so multiply all values of z by 0.5 and t by 5. In Octave:

```
z=0.5*z;
t=5*t;
```

Now we must fit a polynomial to t and a 6th order polynomial usually works well. In Octave:

```
a=polyfit(z,t,6)
```

gives:

```
ans =

 -37.3101e-12
  12.2396e-09
  -1.3825e-06
  58.0277e-06
-827.1508e-06
   1.4997e-03
   5.0112e+00
```

where the first term is the 6th order co-efficient of the polynomial, the penultimate term is the slope – typically entered as an angle so we need to take its arc-tangent – and the last term is the zeroth order, i.e. the wedge thickness. A good approach is to trace rays backwards from the wedge exit surface and place the bevel where they converge, see figure 9. This gives us a good enough starting point to begin optimising with conventional lens design software.

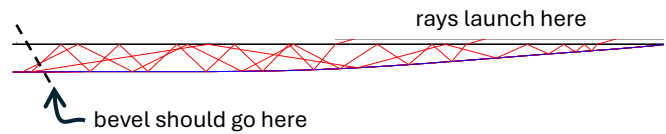

*Figure 9: Ray trace of wedge with polynomial from flatslab.m*

6. Optimisation at finite thickness

Ideally, all rays emerging from one point at the wedge entrance should reach the wedge exit surface at the critical angle, i.e. they should be collimated. To some extent, therefore, the wedge is to act like a lens with the thick end acting as the focal plane of that lens. Within the limit set by figure 8, we want to maximise aperture at that focal plane so our criterion of optimisation is that two ray bundles, each emanating from one end or other of the entrance aperture, should emerge collimated with other rays from the same bundle over the whole exit surface of the wedge. The variables are the coefficients of the polynomial determining the shape of the bottom surface of figure 9.

Optimisation tends to cause the slab to become slightly bulged in the centre which slightly decreases the number of reflections undergone by a ray at the critical angle and therefore shortens the wedge. The reason for the bulge is that its exit slope more closely matches the entry slope of the wedge and we could easily have chosen to introduce such a bulge in the very thin model of which more later.

Classic lens designers are well accustomed to using machine optimisation to improve their designs but it can sometimes be useful to use the stratagem of section 3 to improve designs using conventional ray-tracing through wedges of finite thickness.

Begin by tracing a ray from one corner of the bevel at the critical angle, and a ray at the opposite corner of the bevel at such an angle that both leave the guide at the same point at the beginning of the wedge. Then slightly reduce the angle of both rays versus the guide axis so that the point at which they emerge moves slightly towards the thin end of the wedge. Continue this process until one or other ray undergoes one more or one less bounce than required. At this point, we must either increase local thickness if we want one or other ray to exit with an extra bounce, or reduce thickness if we want the ray to exit with one less bounce.

A simple stratagem is to calculate a range of displacement (z) versus position (y) values for the polynomial in a spreadsheet, slightly alter the displacement at the required location, fit a polynomial to the new set of (y,z) values and enter them by hand into the ray-tracing software. Laborious as this may seem, it is remarkably effective and can be quicker than trying to get optimisation software to do the same job.

7. Curved wedges

The wedge is a light guide and users of optical fibre light guides are used to having them flex: can the wedge do the same? Merely designing a flat light-guide then curving it in software as if it were made of rubber gives poor results. Instead, we need to include the curve in the thin model.

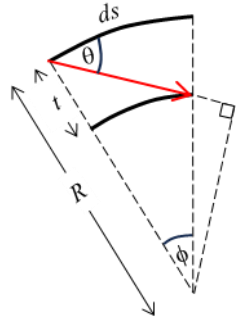

*Figure 10: geometry for a ray propagating along a curved guide*

If a wedge is to be curved, then equation (2) no longer holds and we have to replace it with an expression derived from figure 10.

| | | |
|---|---|---|
| For one bounce of the ray: | $ds = R\phi$ | (8) |
| For $b$ ray bounces: | $ds = R\phi b$ | (9) |
| From right-angled triangle: | $R \cos\theta = (R - t)\cos(\theta - \phi)$ | (10) |
| $\phi$ is arc-angle for 1 bounce: | $\phi = \theta - \cos^{-1}\left[\left(\dfrac{R}{R-t}\right)\cos\theta\right]$ | (11) |
| Number of bounces, $b$: | $b = \dfrac{ds}{R\left(\theta - \cos^{-1}\left[\left(\dfrac{R}{R-t}\right)\cos\theta\right]\right)}$ | (12) |

Note that for this expression to be meaningful, we have to use the (finite) values of $t$ and $R$ that we want for our finished design. Nevertheless, we can continue to treat $b$ as a real number (and not necessarily an integer) and the rule that $t\sin\theta$ is constant still holds so the procedure of figure 5 remains useful as a small-thickness model. Algorithm curvedlslab.m in appendix 2 shows how to calculate a wedge on a curve of constant radius and this was the procedure used to calculate the curved wedge of reference [8]. If the radius varies with distance along the guide, then it is necessary to alter the algorithm so that the shape of the guide can be specified (e.g. with a polynomial) before the wedge is calculated.

8. What limits wedge length?

The wedge tends to be of approximately equal length to the slab - can it be made much longer? Just as figure 8 depicts the zig-zag path of a ray through a wedge as a straight ray through a stack of wedges, we can depict the path of a ray through a slab of constant thickness as a straight ray through a stack of slabs. Figure 11 shows a diagram depicting a ray that passes through a slab then a wedge and we have assumed that both slab and wedge are very thin so that the thick ends of the wedge combine into a curve. In figure 11, the ray exits immediately after entering the wedge. Figure 12, however, traces a ray launched at a shallow angle to the guide axis such that the ray exits near the top of the wedge, and we have rolled the stack of wedges against the end of the slab so that both touch as the ray passes from one stack to the other.

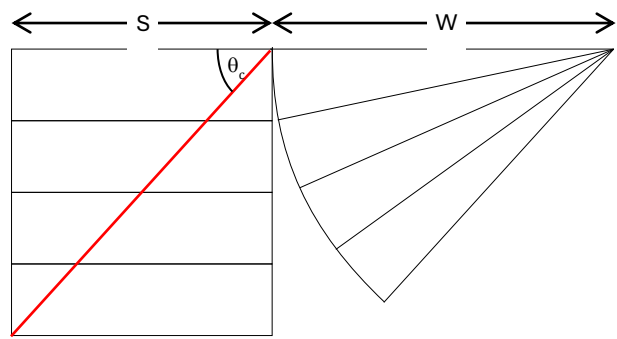

*Figure 11: Diagram for ray launched into the wedge surface near its thick end*

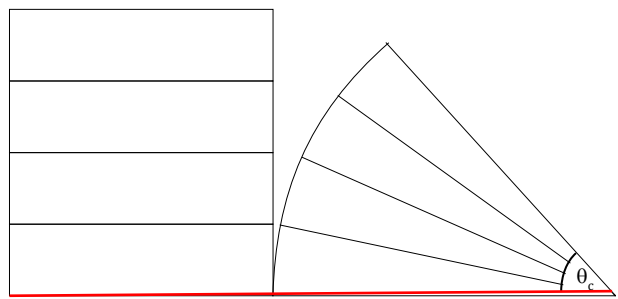

*Figure 12: Diagram for ray launched into the wedge surface near its thin end*

Figure 12 shows that the angle subtended by the stack of wedges is the critical angle relative to the guide axis i.e. $\pi/2 - \sin^{-1}(1/n) = \cos^{-1}(1/n)$. It follows that the length of the arc formed by the thick ends of the wedge stack is $W \cos^{-1}(1/n)$. This must also be the height of the stack of slabs since the stack of wedges rolls against this, so by trigonometry in figure 11:

$$S = W \cos^{-1}(1/n) / \tan \theta_c \qquad (13)$$

$$= W \cos^{-1}(1/n) \times \frac{1}{\sqrt{n^2 - 1}} \qquad (14)$$

We see why the slab is slightly shorter than the wedge at common refractive indices such as 1.5 or 1.6 and that the length of the slab reduces with increasing refractive index, $n$.

It is helpful to imagine a set-up where, just as rays get to the end of the slab, they undergo reflection and the slab magically becomes a flat-sided wedge. If the mirror at the end of the slab is curved with twice the radius of curvature of the stack of wedges, then we have the situation of figures 11 and 12 but we can interpret it by allowing that it is the curved mirror that collimates rays radiating from the stack entrance. In figure 9, a polynomial has been used to smooth out the kink between slab and wedge into a gradual curve and this clearly also has the effect of collimating rays.

If the ideal is that a wedge perfectly collimates rays radiating from its input, how good can that collimation be?

9. How good can collimation get?

Stacks of flat slabs and stacks of flat-sided wedges can be drawn on a flat surface but it is possible also to draw stacks of guides like that of figure 9 provided that the surface on which they are drawn has appropriate curvature. This is perhaps merely a curiosity except in the case where refractive index tends towards infinity (so that the length of the slab tends to zero) where the surface is a sphere and rays travel like yachts on the equator all heading to the north pole.

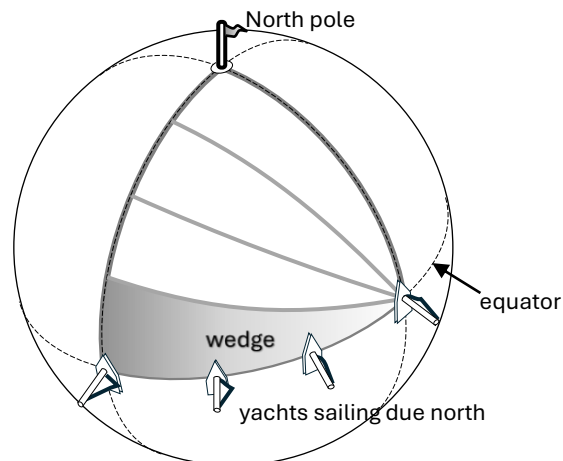

*Figure 13: In a wedge with infinite refractive index, rays injected through the surface converge to a point at the thick end just as yachts heading northwards from the equator converge.*

In this case, it clearly is possible in principle to get perfect focus, the caveat being that in reality, the wedge will have finite thickness so the surface on which the wedge is stacked will rather be a faceted sphere.

We should not assume that perfect focus is only possible with wedges of infinitesimal thickness: rays incident parallel to the focus of a parabola will focus to a point and we can think of a section of a parabola as a wedge in which rays undergo only one reflection before exit.

Our interest is in wedges with the refractive indices of real materials: for these materials, is there an equivalent of the sphere whose shape will allow rays to focus to a point?

Return to our very thin model and consider that the slab need not have constant thickness although it cannot anywhere be thinner than at the point of entry since rays would escape. On the other hand, parts of the slab can be thicker and of course must be if we are to eliminate the kink at the transition between slab and wedge. If we introduce a bulge in the slab in order to eliminate the kink, the number of reflections in the slab reduces so the wedge becomes shorter.

Our ideal is that at the slab to wedge transition there is continuity of slope but also continuity of curvature since our rays must be collimated at exit and a sudden change in curvature will hamper collimation. For the same reason, we would like the rate of change of curvature to be constant throughout the slab and this allows us to define a polynomial for a slab and wedge with optimal collimation.

Let the slab thickness $t(z)$ be a fourth order polynomial and choose its coefficients so that the thickness equals one at slab entrance and slab exit and the slope, curvature and rate of change of curvature of the slab at exit equal those of the wedge at entry:

| | | | |
|---|---|---|---|
| For the slab, let | $t$ | $= Dz^4 + Cz^3 + Bz^2 + Az + 1$ | (15) |
| Slope continuous at z=0: | $t'(z=0)$ | $= A$ | (16) |
| Curvature at z=0: | $t''(z=0)$ | $= 2B$ | (17) |
| Rate of curvature at z=0: | $t'''(z=0)$ | $= 6C$ | (18) |
| Thickness=1 at slab entry, i.e. at z=-$z_0$: | $1$ | $= D(-z_0)^4 + C(-z_0)^3 + B(-z_0)^2 + A(-z_0) + 1$ | (19) |
| | $0$ | $= D(-z_0)^3 + C(-z_0)^2 + B(-z_0) + A$ | (20) |
| | $D$ | $= \dfrac{z_0^2 C - z_0 B + A}{z_0^3}$ | (21) |

We can write the algorithm if Appendix 3 that calculates the wedge for a slab of constant thickness, finds its slope, curvature and rate of change of curvature at the start of the wedge, imposes this on the end of the slab and finds a new slab profile. The algorithm then finds the wedge for this new slab and continues looping until the profile converges as shown in Figures 14 and 15.

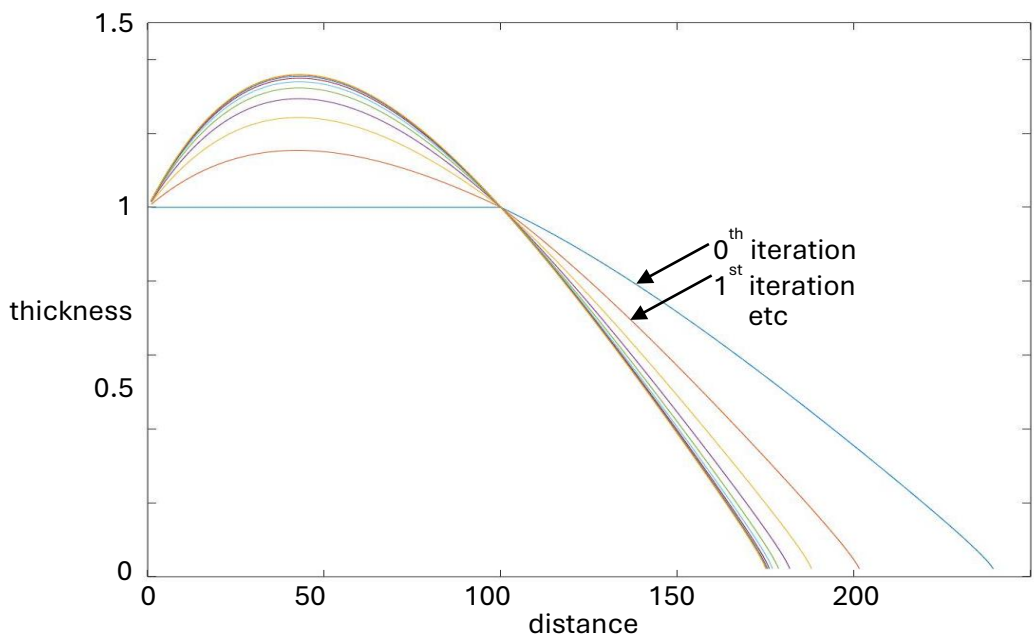

*Figure 14: stages of iteration towards an acrylic guide*

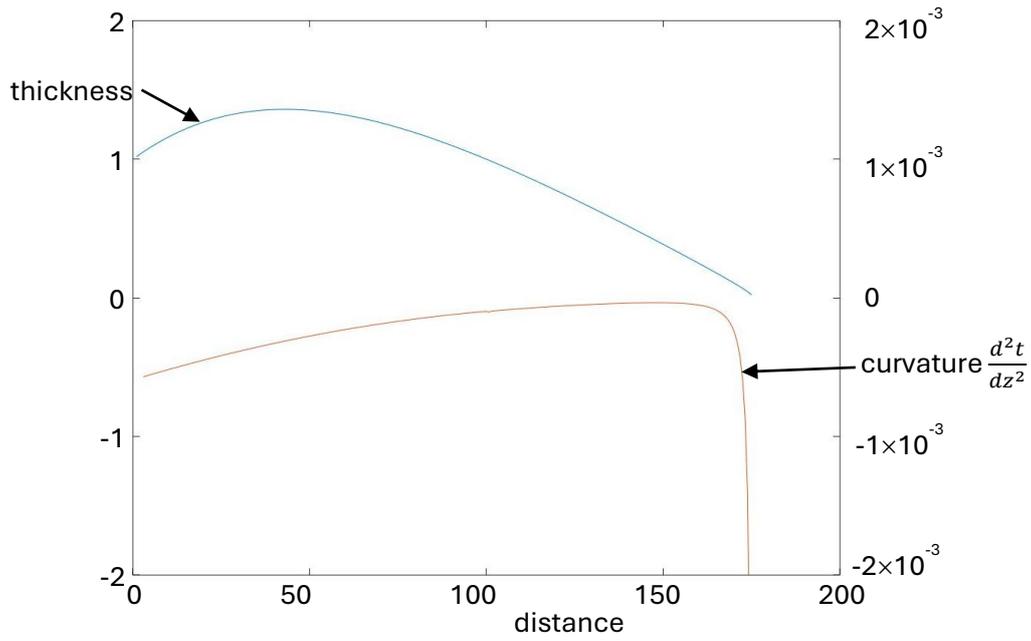

*Figure 15: variation in curvature of the final iteration of an acrylic guide*

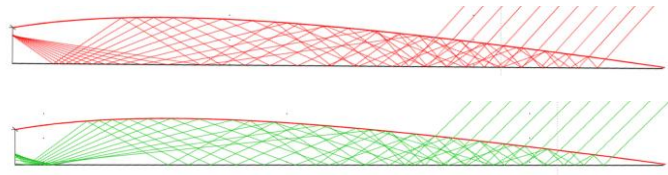

*Figure 16: guide created by algorithm of figure 14 after 5 iterations, thickness scaled by 10. Red rays are launched at critical angle, green rays at critical angle less 1.7°*

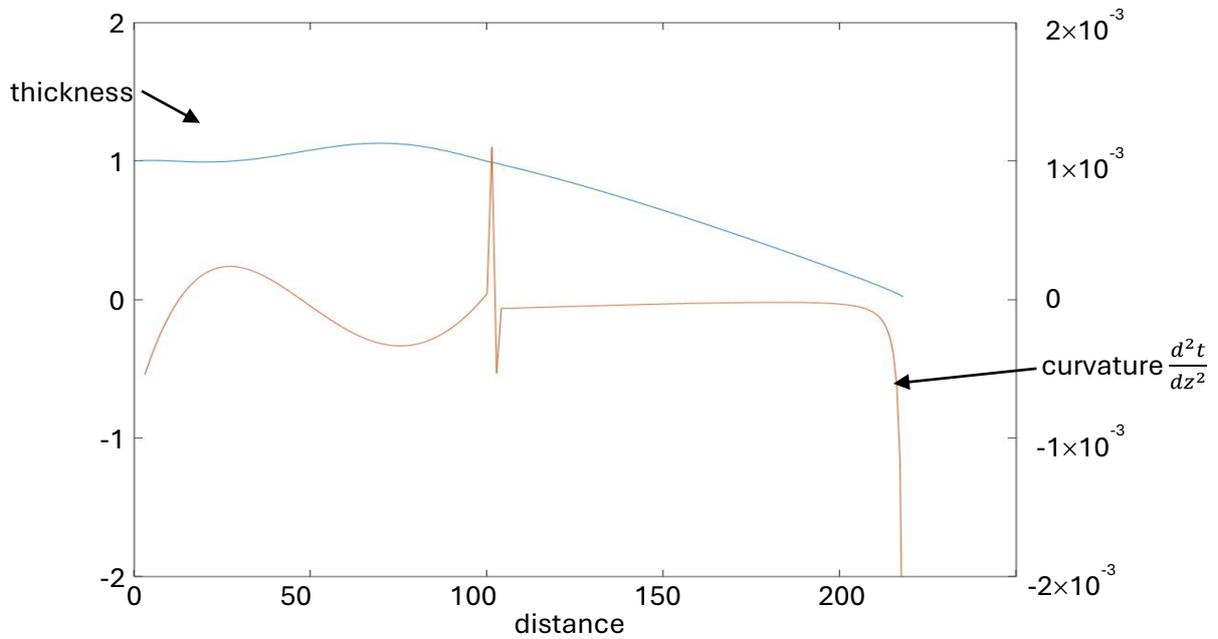

*Figure 17: thickness profile (blue) and curvature(orange) in acrylic where the wedge is 17% longer than the slab*

Note that the guide of figure 16 has been designed entirely with the small thickness model – there has been no machine optimisation. Arguably this design is as smoothly varying as is possible with acrylic because the polynomial of the slab is no more than fourth order. But the slab is longer than the wedge and users often want the slab to be at least shorter than the wedge, not least so that a pair of guides can be placed anti-parallel so that images can be projected throughout the panel. We could make the wedge longer by introducing higher order polynomial terms into the slab at the expense of poorer collimation at exit. In particular, we can force the slope at the entrance of the slab to be zero and this produces a wedge that is approximately 6% longer than the slab. For longer wedges, it is arguably easier to design a slab interactively and Figure 17 shows an acrylic profile where the wedge is 17% longer than the slab.

## 10. Conclusions

The thickness versus distance profile of a thin wedge light-guide can easily be calculated using the principle that thickness multiplied by the sin of ray angle is constant in a smoothly varying guide. Scaling the profile up to larger thicknesses often gives a design good enough that little optimisation is needed.

## 11. Acknowledgements

The author would like to thank Tim Large, Neil Emerton and Liying Chen for their help over the years in developing these ideas.

## Appendix 1

### flatslab.m

```
%***********************************************************************
% I design a wedge with a flat slab.                                   *
%***********************************************************************

clear;
n=1.492;                        %n=refractive index of acrylic.
seg=100;                        %seg=number of segments used to define slab.
z0=100;                         %z0=length of slab

z=[1:seg]*z0/seg;               %z=distance along slab, give z seg segments,
t(1:seg)=1;                     %t=thickness  of slab, make t uniform.
[t,z,b0]=getwedge(t,z,n);       %Calculate the wedge

plot(z,t)
```

### Function getwedge.m

```
%* getwedge takes the thickness versus distance (t,z) co-ordinates of the   *
%* slab and the refractive index, n, and adds the co-ordinates of the wedge *
%* that, in combination with the slab, gives a constant number of bounces,b0 *

function [t,z,b0] = getwedge (t,z,n)
seg=length(z);
t0=t(seg);                      %t0=thickness at z=z0.
z0=z(seg);                      %z0=point where slab ends, wedge starts.

dz(1)=z(1);                     %dz=length of each segment.
dz(2:numel(z))=z(2:numel(z))-z(1:numel(z)-1);

qc=pi/2-asin(1/n);              %Find qc, critical angle & choose some values for
q0=[qc-pi/360:-pi/360:pi/360];%q0=angle vs guide tangent at z0 (wedge entry).

tc=t0*sin(q0)/sin(qc);          %tc=thicknesses at which rays exit, by Lagrange.

%First count the number of bounces of the ray that exits at z=z0 (wedge entry).
q=asin((t0./t)*sin(qc));        %q =  angle in each segment of the ray,
db=dz.*tan(q)./(2*t);           %db=bounces in each segment of the ray and
b0=sum(db);                     %b0 = total bounces in slab of the ray.

%Now arrange that all other rays have the same number of bounces as the first.
c=1;                            %Find dz so all other rays do b0 bounces.
while c<=numel(q0);
  q=asin((t0./t)*sin(q0(c))); %q =  angle in each segment of ray c.
  db=dz.*tan(q)./(2*t);       %db=bounces in each segment of ray c.
  bb=sum(db);                 %bb=sum of bounces in guide to penultimate segment
  dz(seg+c)=(b0-bb)*2*tc(c)/tan(qc); %Add a segment to the wedge so that the
  t(seg+c)=tc(c);             %total number of bounces of ray c at exit=b0.
  z(seg+c)=dz(seg+c)+z(seg+c-1); %Update z.
  c=c+1;                      %Prepare to trace next ray.
end
endfunction
```

## Appendix 2
## curvedslab.m

```
%***********************************************************************
% I design a curved wedge with a bulging slab.                         *
% Adrian Travis 10 July 2025                                           *
%***********************************************************************

clear; clf;
n=1.585;         %n=refractive index.
seg=100;         %seg=number of slab segments.
R=82;            %R=outer radius of curvature.
s0=70;           %s0=length of slab

t0=5.0357;       %choose slab thickness,
a1=-0.0072136;   %slope,
a2=0.0006968;    %curvature and
a3=-8.5629e-6;   %rate of change of curvature

s=(s0/seg)*[1:seg]; %s=distance along guide.
t= t0+a1*s+a2*s.^2+a3*s.^3; %Slab thickness is a cubic

[t,s,b0]=getcurvedwedge(t,s,n,R);

plot(s,t);       %Plot thickness versus length along the arc.

%We have now calculated wedge thickness versus length, but we need to bend it
%so that it has radius of curvature R and we need to find its y,z co-ordinates.
%We center the wedge on s0 and the profile will be on the inner surface.

y   =  (R-t).*sin((s-s0)/R); %y co-ord of inner surface of wedge and
zin =R-(R-t).*cos((s-s0)/R); %z co-ord of inner surface of wedge.
zout=[y.*y/R]./[1+sqrt(1-y.*y/R^2)]; %z co-ord of outer surface of wedge.
figure 2; plot(y,zout,y,zin); axis("equal"); %Draw the curved wedge.

%Lastly, we need to fit a polynomial to the inner surface so that it can be
%expressed in ray-tracing software.

zc =t(seg)+[y.*y/(R-t0)]./[1+sqrt(1-y.*y/(R-t0)^2)]; %zc=inner arc, radius R-t0.
poly=zin-zc;     %Subtract the inner arc from the wedge surface. Fit a polynomial
a=polyfit(y,poly,6); %so that we can express it in ray-tracing software.
disp('wedge polynomial, coefficients 6 to 0:')
format shorteng
disp(a')
```

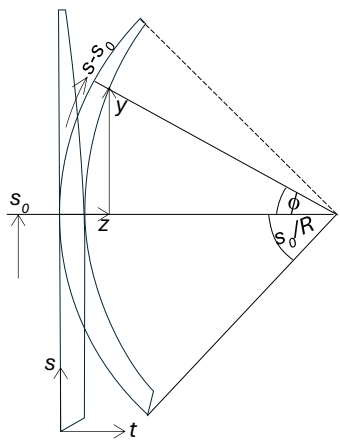

*Figure showing geometry of procedure by which wedge is curved*

# getcurvedwedge.m

```
## Copyright (C) 2025 arlt1
##
## This program is free software: you can redistribute it and/or modify it
## under the terms of the GNU General Public License as published by
## the Free Software Foundation, either version 3 of the License, or
## (at your option) any later version.
##
## This program is distributed in the hope that it will be useful, but
## WITHOUT ANY WARRANTY; without even the implied warranty of
## MERCHANTABILITY or FITNESS FOR A PARTICULAR PURPOSE.  See the
## GNU General Public License for more details.
##
## You should have received a copy of the GNU General Public License
## along with this program.  If not, see
## <https://www.gnu.org/licenses/>.

## Author: Adrian Travis
## Created: 2025-07-10

% getcurvedwedge takes the thickness versus length (t,s) co-ordinates *
% of the slab, its refractive index, n, & its radius of curvature, R   *
% and adds the co-ordinates of the wedge with that same radius of      *
% curvature that, in combination with the slab, gives a constant       *
% number of bounces, b0.                                               *

function [t,s,b0] = getcurvedwedge (t,s,n,R)
seg=length(s);
t0=t(seg);                   %t0=thickness at s=s0.
s0=s(seg);                   %s0=point where slab ends, wedge starts.

ds(1)=s(1);                              %ds=length of each segment.
ds(2:numel(s))=s(2:numel(s))-s(1:numel(s)-1);

qx=acos(R/(R+t0));           %qx=max value of q round curve.
qc=pi/2-asin(1/n);           %Find qc, critical angle and
q0=[qc-pi/36:-pi/36:qx+pi/36]; %q0=launch angle vs guide tangent.
tc=t0*sin(q0)/sin(qc);       %Lagrange gives t at exit.

%First count the number of bounces of the ray that exits at s=s0 (wedge entry).
q=asin((t0./t)*sin(qc));     %q=ray angle in each segment
db=(ds/R)./(q-acos((R./(R-t)).*cos(q))); %db=bounces/segment
b0=sum(db);                  %b0=total bounces in slab

%Now do the same for each ray & add a bit of wedge so it exits after b0 bounces.
c=1;                         %Set up counter.
while c<=numel(q0);          %For each ray (until there are none left)...
  q=asin((t0./t)*sin(q0(c))); %Find the ray angle in each existing segment.
  db=(ds/R)./(q-acos((R./(R-t)).*cos(q))); %db=bounces/segment
  bb=sum(db);                %bb=bounces in guide so far. Add section
  t(seg+c)=tc(c);            %so total bounces of ray at exit=b0.
  ds(seg+c)=(b0-bb)*R*(qc-acos((R/(R-t(seg+c)))*cos(qc))); %Find segment length,
  s(seg+c)=ds(seg+c)+s(seg+c-1); %and update s.
  c=c+1;                     %Increment the counter
end
```

# Appendix 3
# bestcollimation.m

```matlab
%**********************************************************************
% I find a quartic polynomial for a slab with a smooth, flat gapless  *
% wedge. Slope and curvature are both continuous. This creates figure *
% 15 of 'Wedge design.doc'.                                           *
%                                     Adrian Travis 13 July 2025      *
%**********************************************************************

clear;
n=1.492;                %n=refractive index (here, I choose acrylic).
seg=100;                %seg=number of segments that are to constitute slab.
z0=100;                 %z0=length of slab

t0=1;                   %t0=thickness at z=0, t=thickness at z>0.
slabg=[0, 0, 0, 0, t0]; %slabg=4th order polynomial slab thickness.Start flat.

for ff=1:5;             %5 loops are enough to reach optimum.
  z=[1:seg]*z0/seg;     %z=distance along flat guide, give z seg segments,
  t=polyval(slabg, z-z0); %& use the curve to get a thickness at each value of z
  [t,z,b0]=getwedge(t,z,n); %then calculate the wedge.

  %Now prepare to plot slab+wedge thickness, slope and curvature...
  dz=0;                 %Reset dz.
  dz(1)=z(1);                   %dz=length of each segment.
  dz(2:numel(z))=z(2:numel(z))-z(1:numel(z)-1);

  dt = (t(2:numel(t))   -t(1:numel(t)-1))  ./dz(2:numel(t)); %Find slope,
  dt2= (dt(2:numel(dt)) -dt(1:numel(dt)-1)) ./dz(3:numel(t)); %curvature &
  dt3=(dt2(2:numel(dt2))-dt2(1:numel(dt2)-1))./dz(4:numel(t)); %d(curvature)/dz.

  plot(z,t,z(3:numel(z)),1000*dt2); %plot thickness & curvature
  ylim([-2*t0,2*t0]);    %This fixes the axes so as to cut out any spike.
  pause(.4);             %Give the user time to observe the latest iteration.

  %Now, update the slab.
  A=dt(seg);             %A is slope at slab/wedge transition,
  B=dt2(seg)/2;          %B is curvature (note division by 2!) and
  C=dt3(seg+1)/6;        %C is rate of change of curvature for t=0 at z=0.
  F=(z0^2*C-z0*B+A)/z0^3; %F is rate of change of rate of change of curvature.
  slabg=[F, C, B, A, t0]; %Enter new slab shape and go back to recalculate wedge.
end

%Now fit a curve to slab+wedge
m= polyfit(z,t,6);       %Fit a 6th order polynomial to both slab and wedge.
tt=polyval(m,z);         %tt=points on polynomial
figure 2                 %In a second window..
plot(z,t,z,tt,z,20*(tt-t)); %.. plot slab+wedge, polynomial & the difference.
legend('design','fitted polynomial','error')
title('fitting polynomial'); xlabel('distance'); ylabel('thickness');
set (gca, "FontSize", 15);  set (legend, "FontSize", 10);
```